\documentclass[prl,twocolumn,showpacs,floats,superscriptaddress]{revtex4}

\newcommand{\sio}{SiO$_{2}$ }

\newcommand{\Pb}{$\mathrm{P}_{\mathrm{b}}$ }
\newcommand{\Pbm}{$\mathrm{P}^{-}_{\mathrm{b}}$ }

\usepackage{times}
\usepackage{mathptmx}
\usepackage{graphicx}
\usepackage{dcolumn}
\usepackage{bm}
\bibpunct{}{}{,}{s}{}{}
\begin{document}
\preprint{PRL}

\title{Spin--dependent processes at the crystalline Si-SiO$_2$ interface at high magnetic fields}

\author{D. R. McCamey}~\email{dane.mccamey@physics.utah.edu}
\affiliation{Department of Physics, University of Utah, 115 South
1400 East Rm 201, Salt Lake City, Utah 84112}
\author{G. W. Morley}
\affiliation{London Centre for Nanotechnology and Department of
Physics and Astronomy, 17-19 Gower Street, London WC1H 0AH, United
Kingdom }
\author{H. A. Seipel}
\affiliation{Department of Physics, University of Utah, 115 South
1400 East Rm 201, Salt Lake City, Utah 84112}
\author{L. C. Brunel}
\affiliation{Center for Interdisciplinary Magnetic Resonance,
National High Magnetic Field Laboratory at Florida State University,
Tallahassee, Florida 32310, USA}
\author{J. van Tol}
\affiliation{Center for Interdisciplinary Magnetic Resonance,
National High Magnetic Field Laboratory at Florida State University,
Tallahassee, Florida 32310, USA}
\author{C. Boehme}~\email{boehme@physics.utah.edu}
\affiliation{Department of Physics, University of Utah, 115 South
1400 East Rm 201, Salt Lake City, Utah 84112}

\date{\today}


\begin{abstract}
An experimental study on the nature of spin--dependent excess charge
carrier transitions at the interface between (111) oriented
phosphorous doped ([P]$\approx 10^{15} \mathrm{cm}^{-3}$)
crystalline silicon and silicon dioxide at high magnetic field
($B_0\approx 8.5\mathrm{T}$) is presented. Electrically detected
magnetic resonance (EDMR) spectra of the hyperfine split $^{31}$P
donor electron transitions and paramagnetic interface defects were
conducted at temperatures in the range 3 K$\leq T\leq$12 K. The
results at these previously unattained (for EDMR) magnetic field
strengths reveal the dominance of spin--dependent processes that
differ from the previously well investigated recombination between
the $^{31}$P donor and the $\mathrm{P_b}$ state, which dominates at
low magnetic fields. While magnetic resonant current responses due
to $^{31}$P and $\mathrm{P_b}$ states are still present, they do not
correlate and only the $\mathrm{P_b}$ contribution can be associated
with an interface process due to spin--dependent tunneling between
energetically and physically adjacent $\mathrm{P_b}$ states. This
work provides an experimental demonstration of spin-dependent
tunneling between physically adjacent and identical electronic
states as proposed by Kane for readout of donor qubits.
\end{abstract}


\maketitle


Phosphorus doped crystalline silicon (c-Si:P) is one of the most
widely utilized semiconductor materials, with applications ranging
from conventional microelectronics\cite{sze} to proposed and
presently widely investigated concepts for spintronics
\cite{Appelbaum2007} and spin-based quantum information processing
(QIP)\cite{kane}. Silicon based spin--QIP and spintronics concepts
aim to utilize the comparatively weak spin--orbit coupling present
in this material, and the correspondingly very long spin--coherence
times~\cite{kane,Appelbaum2007}, as well as the impact of
spin--selection rules on electronic transitions which can be used
for spin readout~\cite{Boe5}. Most of these applications involve
electrical transport and spin manipulation at or near the
silicon--silicon dioxide (SiO$_2$) interface, making the
understanding of spin processes in this region extremely important.
Numerous studies of spin--dependent transport and recombination at
the interface between c-Si:P and SiO$_2$ have recently been
undertaken, with the aim of identifying and understanding these
mechanisms~\cite{Lenahan1998,Friedrich2005,stegner06}, and showing
that they can be utilized for the observation of very small
ensembles of donors\cite{McCamey2006} and coherent spin
motion~\cite{stegner06,Huebl2007}. Additionally, spin dependent
transport in two dimensional electron gases at the c-Si/SiO$_2$
interface has been demonstrated~\cite{Ghosh1992,Lo2007}. However, no
systematic study of such processes at high magnetic field has been
conducted to date, with the only data at magnetic fields $B_0>400$
mT given by a single electrically detected magnetic resonance (EDMR)
spectrum recorded at $B_0=7.1$ T and a temperature $T=4$ K
~\cite{Honig1978}.

In the following, a systematic investigation of the spin dependent
processes at the interface between c-Si:P and SiO$_2$ are presented
for high magnetic fields ($B_0\approx8.5$ T) at temperatures in the
range 3 K$\leq T\leq$12 K.We show that the dominant spin-dependent
recombination mechanism at low magnetic fields, recombination
between $^{31}$P and \Pb centers, is not seen at high fields.
Instead, transitions involving only $^{31}$P donors or
$\mathrm{P_b}$ centers dominate the observed EDMR signals. This
study focuses in particular on the nature of the $\mathrm{P_b}$ only
transition which has previously been observed at low magnetic fields
and nominally undoped c-Si--SiO$_2$ interfaces \cite{Friedrich2005},
but has, however, not been observed in the presence of $^{31}$P
donors.

\begin{figure}
\centering\includegraphics[width=7cm]{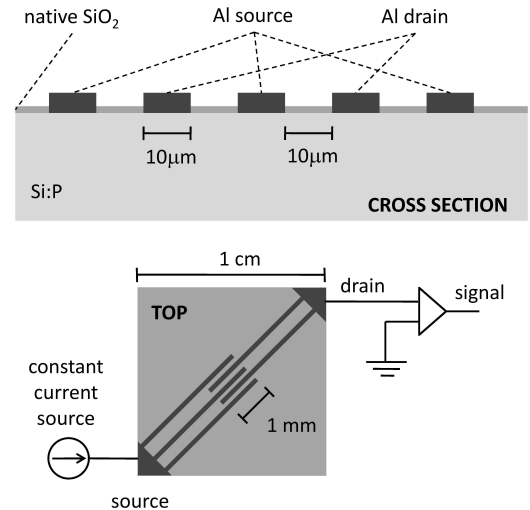}
\caption{\label{fig:device}A sketch of the sample used in these
experiments. The sample is fabricated on Si(111) doped with $10^15$
phosphorus donors/cm$^3$. contacts are made to the sample using
aluminum contacts, and the surface is covered with a native \sio
layer. A simple measurement circuit is also shown. The sketch is not
to scale, although the indicated dimensions are accurate.}
\end{figure}

Experimentally, we used prime grade Cz--grown c-Si(111) with a
phosphorus donor concentration [P]$\approx 10^{15}\mathrm{cm}^{-3}$.
The sample was contacted by thermal evaporation of a 100 nm Al-layer
after a surface clean and subsequent removal of the native oxide by
wet treatment with hydrofluoric acid. Following this procedure and
the structuring of the sample contacts by a photolithographic
lift--off procedure, a native \sio layer was formed on the surface
between the contacts due to the exposure of the sample to air at
room temperature. Similarly to previous studies of spin--dependent
recombination and transport at low magnetic fields
\cite{Lepine72,Thornton1973,Kaplan1978}, we used EDMR to investigate
these processes. With this technique, the photocurrent through a
sample is monitored while electron spin resonance is used to
manipulate the spin of paramagnetic centers involved in
spin--dependent transitions. The latter are detected by measurement
of currents which change from a constant offset value under spin
resonance conditions~\cite{stutz}. In order to perform spin
resonance at $B_0\approx 8.5$ T, the quasi optical 240 GHz
heterodyne spectrometer facility of the National High Magnetic Field
Laboratory in Tallahassee, Florida was used~\cite{Tol2005}. A sample
compatible to the geometric constraints of the
Fabry--P$\acute{\mathrm{e}}$rot (FP) resonator of the spectrometer
was used for the experiments, and is shown schematically in figure
\ref{fig:device}. It consists of an approximately 330 $\mu$m thick,
8 mm x 8 mm silicon substrate sandwiched between two 160 $\mu$m
thick quartz slabs needed as antireflection coatings to allow the
240 GHz radiation to be coupled into the silicon bulk. The
electrical contacts to the device are a 100 $\mu$m wide grid
structure consisting of five 10 $\mu$m wide interdigitated fingers,
with 10 $\mu$m separation between opposite fingers. The contact
fingers were approximately 6 mm long, yet fingers belonging to the
two opposite contacts overlapped by only 1mm. This geometry ensured
that (i) the active region of the sample was located on the optical
axis of the FP resonator such that the microwave field $B_1$ was
maximal and homogeneous throughout the active area; (ii) the
external contacts of the sample, which were contacted with silver
paste, were well outside the $B_1$ field such that they could not
distort the FP resonator modes; and (iii) due to the length of the
contacts (almost 12 mm, stretching across the entire beam diameter),
all metal structures within the beam diameter were aligned
perpendicular to the polarization of the $B_1$ field, reducing loss
due to microwave absorption. The high magnetic field, $B_0$, is
aligned normal to the sample surface. The photocurrent needed for
the EDMR experiments was induced by white light (cold light)
generated by a xenon discharge lamp, filtered of its infrared
component and coupled into the sample via an optical fiber.

\begin{figure}
\centering\includegraphics[width=9cm]{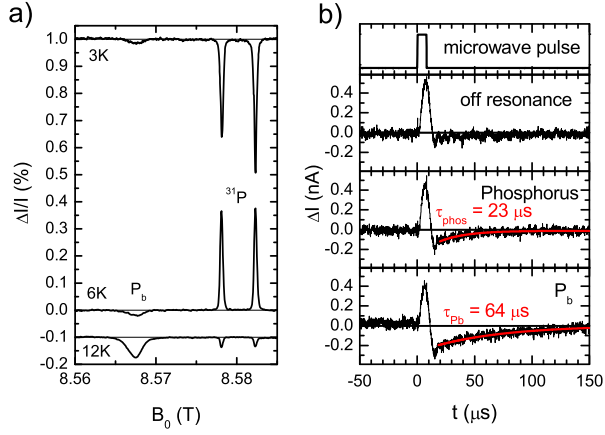}
\caption{\label{fig:12KSpectraandTimeTraces}a) Plots of the relative
photocurrent changes as a function of the applied magnetic field,
$B_0$ at temperatures $T=3$K, 6K and 12K. The data was taken with a
photocurrent $I = 600$ nA. The signal was obtained by microwave
chopping ($500\mu$s pulse length, 1 kHz shot repetition rate) and
lock--in detection of the photocurrent. b) Plots of the photocurrent
changes as a function of time following a microwave pulse with
$\tau_{p} = 8\mu$s length for magnetic fields off--resonance,
on--resonance with the low field $^{31}$P peak, and on--resonance
with the \Pb peak, at $T=12$ K. The data was collected from the
average of many transients measured with a 1 kHz shot repetition
rate.}
\end{figure}

For the data acquisition, the microwave radiation was modulated
which allowed a lock--in detection of the magnetic resonance induced
current changes. The relative current change, $\Delta I/I$, observed
is shown in Fig. \ref{fig:12KSpectraandTimeTraces}a for $T=$3 K, 6 K
and 12 K. There are three resonances clearly visible which (as for
all data presented in this study) were fit by Gaussian functions.
The two resonances at the highest magnetic fields are separated by
$4.2$ mT, as expected from phosphorus donor electrons due to their
hyperfine coupling to the donor nuclear spin. These resonances were
used to calibrate the magnetic field axis (with the applied
frequency set to $240$ GHz) due to the drift in the superconducting
magnet and the internal field due to the polarized
electrons.\footnote{The offset never exceeded 10 mT, and was usually
less than 4 mT. We assume that the \Pb defects feel the same
internal field as the phosphorus donors.} The resonance at lower
magnetic field, at a g-factor of $g=2.0014$, is assigned to the \Pb
interface defect (a silicon dangling bond)\cite{Lenahan1998} due to
the agreement between the experimentally determined $g$-factor and
the accepted literature value of  $g=2.0014$, for $B_0$ parallel to
the $\langle 111 \rangle$ direction. In contrast to experiments at
lower magnetic fields, the \Pb resonance here is well separated from
the two phosphorus resonances, outside the field range that connects
the two hyperfine peaks. This is expected as while the magnetic
field separation of the \Pb and phosphorus resonances depends on the
$g$-factor difference, the phosphorus hyperfine splitting is
constant ($4.2$ mT) for high magnetic fields ($B>>4.2$ mT).

The data show that the peak intensity, $A$, (defined as the
integrated resonance lines = areas of the Gaussian fits) of the \Pb
resonance and the sum of the two hyperfine coupled $^{31}$P
resonances have no correlation. While at $T=12$ K, the \Pb resonance
is $\approx 6$ times larger than the sum of the areas of the two
$^{31}$P resonances, it is smaller at $T=6$ K and $T=3$ K. Moreover,
at $T=6$ K, the signs of the $^{31}$P resonances are positive, in
contrast to the sign of the \Pb resonance, which is consistently
negative at all measured temperatures. These observations are in
stark contrast to low field EDMR, where the dominance of the
$^{31}$P-\Pb pair mechanism causes a complete correlation between
the intensities of \Pb and the intensity sum of the two $^{31}$P
peaks~\cite{stegner06}. Additional evidence that different
spin--dependent processes dominate at high magnetic fields is given
by the current transient following a single, short microwave pulse.
Fig.~\ref{fig:12KSpectraandTimeTraces}b) shows such transients
off--resonance, and on--resonance for the phosphorus and the \Pb
peaks. These transients were taken with a pulse length of 8 $\mu$s.
A short microwave induced current which decays after approximately
$t=30$ $\mu$s is seen in all traces. Additionally, a decrease in the
current ($\Delta I < 0$) is seen for both the phosphorus and \Pb
resonances. The return of the current to the steady-state value can
in both cases be fit with a simple exponential decay. This is
different to the more complex double exponential quenching/
enhancement expected for the $\mathrm{^{31}P}-$\Pb pairs that are
visible at low magnetic fields. Additionally, the time constants of
the two exponential recoveries, $\tau_{\mathrm{P}_{\mathrm{b}}} =
64(2)$ $\mu$s and $\tau_{\mathrm{phos}} = 23(2)$ $\mu$s for the \Pb
and phosphorus respectively, are very different, further indicating
that the two resonances are not due to the same processes. It shall
be noted that the data in Fig.~\ref{fig:12KSpectraandTimeTraces}b)
shows that EDMR at $B_0\approx 8.5$ T leads to a significantly
smaller ratio between microwave induced artifact currents and
spin--dependent currents than that seen at low fields. This makes
high field EDMR on silicon significantly more sensitive than X-band
EDMR typically conducted at $B_0\approx 340$ mT.

\begin{figure}
\centering\includegraphics[width=70mm]{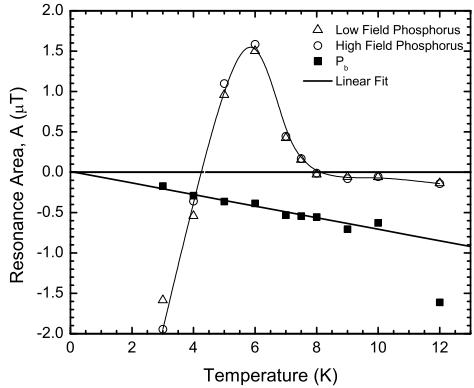}
\caption{\label{fig:Magnitudes_Complex} Plot of the integrated area
$A$ under each of the three resonances as a function of the
temperature. The linear fit to the \Pb data between 3 K and 10 K
shows excellent agreement. The data at 12 K was taken at a different
illumination intensity. The line joining the data for the $^{31}$P
resonances is a guide to the eye.}
\end{figure}

In addition to the spectra displayed in
Fig.~\ref{fig:12KSpectraandTimeTraces}, we have measured EDMR for a
number of other temperatures between 3 K and 10 K. Figure
\ref{fig:Magnitudes_Complex} shows the value of the peak areas of
each of the three resonance peaks, namely the \Pb and both the high
field and low field phosphorus. In order to obtain the maximum area
each peak was fit after correction of the lock-in phase due to the
different dynamic behaviors of the \Pb and $^{31}$P signals. The
magnitudes of the two phosphorus resonances are very similar, as
expected due to the negligible nuclear polarization. However, as
forshadowed in Fig.~\ref{fig:12KSpectraandTimeTraces}, the form of
the temperature dependence is unexpectedly different from low field
EDMR experiments. From high to low temperature, the $^{31}$P signals
are initially negative and smaller than the \Pb signal, becoming
larger and positive between $T\approx 8$ K and $T\approx 4.5$ K,
before again becoming negative, but with a larger magnitude, for
temperatures below $T \approx 4.5$ K. We note that the sign and
amplitude of the resonance at the lowest temperature recorded agrees
with the single spectra reported by Honig and
Moroz~\cite{Honig1978}.

The experimental data presented allows us to exclude a number of
mechanisms as the source for the observed signals and temperature
dependencies. First, the signals observed are probably not
bolometric effects due to resonant heating since this is expected to
exhibit non--linear monotonically decreasing temperature
dependencies which for both the $^{31}$P and the \Pb signals are not
in agreement with the observed conductivity changes. It is possible
that the low temperature ($T<6$ K) $^{31}$P signal has some
bolometric component, however, Honig and Moroz~\cite{Honig1978} have
assigned this to a spin--dependent neutral donor capture and
reemission process, based on a spectra quantitatively similar to
that presented here. Hence, we conclude that the signals observed
must be due to spin--dependent electronic transport or recombination
processes (except for $^{31}$P in the range $T<6$ K). Second, from
the different magnitudes, the different signs and the different
temperature dependencies of the $^{31}$P and the \Pb signals we
conclude that the $\mathrm{^{31}P}-$\Pb interface recombination
mechanism that dominates spin--dependent recombination rates at low
magnetic fields is not responsible for the EDMR signals at high
magnetic fields. Thus, the observed $^{31}$P and the \Pb resonances
must be due to independent electronic processes.  The $^{31}$P
enhancement signal for 4.5 K$<T<8$ K is not understood at this time
in absence of a theoretical framework describing this temperature
behavior. This signal can not be attributed to bolometric effects
for the reasons stated above and because of its sign, as resonant
sample heating is expected to cause a decrease of the conductivity.
Thus, the source of the $^{31}$P enhancement signal can not be
attributed to an interface effect and consequently the only signal
that is clearly due to an interface process is the
$\mathrm{P_b}$--only transition.

\begin{figure}
\centering\includegraphics[width=100mm]{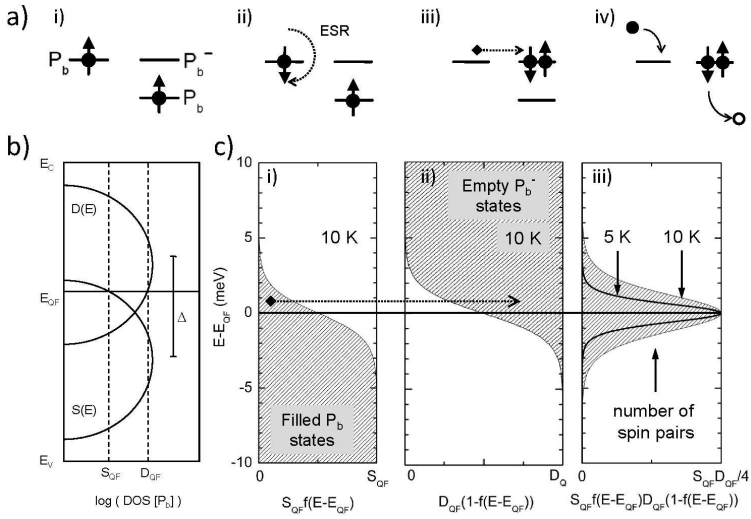}
\caption{\label{fig:model}a) Cartoon of spin--dependent tunneling
between two adjacent uncharged \Pb centers where the ground state
energy of one \Pb matches the \Pbm/\Pb charging energy of the other.
(i) Initially, the pair is in a triplet state and tunneling is not
possible. (ii) Once the spin of one \Pb state is changed the pair
has singlet content and tunneling (iii) is allowed. (iv) An excess
charge carrier pair recombines as it discharges the two charged \Pb
states. b) Sketch of the DOS of \Pb centers within the c-Si
bandgap~\cite{Lenahan1998}. Note that the DOS of \Pb and \Pbm is
assumed to be identical but shifted by the correlation energy
$\Delta$. c) Plot, for $T=10$ K, of the number of (i) filled \Pb
states, (ii) filled \Pbm states and (iii) tunneling pairs as a
function of energy about the quasi-Fermi level $E_{QF}$. The sharper
pair distribution for $T=5$ K is indicated by the solid line. The
plot illustrates that there are fewer pairs of adjacent \Pb and \Pbm
with matching energies when $T$ decreases.}
\end{figure}

We now consider the underlying process leading to the \Pb resonance.
As the \Pb center is a paramagnetic deep interface state
\cite{Lenahan1998}, spin dependent electronic transitions are
described by a two spin-$1/2$ pair model \cite{Lepine72,Kaplan1978}.
Thus, the EDMR signal from the $\mathrm{P_b}$--only transition
 can be due to: (i) Strongly coupled electron pairs, such as the
charged excited state $\mathrm{P_b^{-*}}$, that decay
spin--dependently into a charged ground state $\mathrm{P_b^{-}}$
(this model was first suggested by Friedrich \emph{et
al.}~\cite{Friedrich2005} for the low field $\mathrm{P_b}$--only
signal observed at the interface of intrinsic c-Si to SiO$_2$), (ii)
tunneling or (iii) energy loss hopping between adjacent singly
occupied $\mathrm{P_b}$ ground states with identical or different
energies, respectively. Fig. \ref{fig:Magnitudes_Complex} shows that
the temperature dependence of the $\mathrm{P_b}$--only signal can be
well fit with a linear function without offset (at $T=0$ the EDMR
signal $\Delta I=0$) up to $T\approx 10$ K. This behavior clearly
contradicts (i) as the EDMR signal is not expected to vanish for
small $T$ in this model. Models (ii) and (iii) involve transitions
between adjacent $\mathrm{P_b}$-centers as illustrated for the
tunneling case (model (ii)) in Fig. \ref{fig:model}a). Using
Simmons--Taylor statistics \cite{Simmons1971}(an extension of
Shockley--Read statistics to an arbitrary defect distribution of
states) it can be shown that at finite temperature and under
illumination, the occupancy of the \Pb defects is given by a Fermi
distribution $f(E)=\left(1+\exp\left[-\frac{(E-E_{QF})}{k_B
T}\right]\right)^{-1}$, about a quasi-Fermi energy $E_{QF}$ with
$k_B$ the Boltzmann constant. The density of filled \Pb states close
to the quasi-Fermi energy is thus the \Pb density of states (DOS),
$S(E)$, times the fermi distribution, $S(E)f(E)$ as plotted in
Fig.~\ref{fig:model}c)(i) for $T=10$ K. Similarly, the density of
unfilled \Pbm states is given by $D(E)(1-f(E))$ and plotted in
Fig.~\ref{fig:model}c)(ii) for $T=10$ K, where $D(E)$ is the \Pbm
DOS. The \Pbm DOS is identical to the \Pb DOS except that it is
offset along the energy axis by the positive correlation energy
$\Delta$ associated with double occupancy of the
defect~\cite{Lenahan1998} (i.e., $D(E) = S(E-\Delta)$) as is
illustrated by the sketch of the DOS in the c-Si bandgap shown in
Fig.~\ref{fig:model}b). We note that for the low temperature range
investigated here, the DOS is effectively constant (i.e.,
$S(E_{QF})=S_{QF}$ and $D(E_{QF})=D_{QF}$) as the thermal energy is
small ($k_B T = 0.86$ meV at 10 K) compared with the energy scale
over which the DOS vary. Next, we assume that every \Pb state
interacts with \Pbm states within some interaction radius, $r$,
which is independent of temperature. If we now consider only spin
pairs whose energy levels are aligned (as expected for the tunneling
model (ii)), we obtain a density $n_{p}$, given by
\begin{equation}
n_{p} =\pi r^2S_{QF}D_{QF} \int^{\infty}_{-\infty}
f(E)(1-f(E))dE=\xi T\label{eqn:linearT}
\end{equation}
with the constant $\xi=\pi r^2S_{QF}D_{QF}k_B$. From
Eq.~\ref{eqn:linearT}, we see that the density of spin pairs is
linear in $T$, with no pairs at $T=0$ K. As we anticipate a
proportionality between the number of spin pairs and the EDMR
signal, the energy-conserving model of tunneling between \Pb pairs
is in agreement with the observed temperature dependence of the the
EDMR signals. When we consider energy loss hopping transitions
(model (iii)), the spin pair density becomes
\begin{equation}
n_{p}=\pi r^2\int^{\infty}_{-\infty}S_{QF}f(E)\int_{-\infty}^{E}
D_{QF}(1-f(E'))dE'dE\propto T^2
\end{equation}
in contrast to the experimental results.

We note that the transition from a $\mathrm{^{31}P}-$\Pb
process~\cite{stegner06} at low fields to a \Pb only mechanism at
high fields may be explained by considering the underlying spin
dynamics. The strength of an EDMR signal becomes weaker as the ratio
$\xi=\frac{\Delta\omega}{\gamma B_1}$ of the difference of the
Larmor frequencies in a pair,  $\Delta\omega$, to $\gamma B_1$ drops
below 1, with $\gamma$ being the gyromagnetic ratio~\cite{Raj06}.
Hence, at low fields EDMR signals are dominated by maximized
$\mathrm{^{31}P}-$\Pb signals as $\xi>1$ while $\xi\ll 1$ for the
\Pb pairs with $\Delta\omega\ll\gamma B_1$ due to the almost
identical Land\'e--factors of two \Pb centers. At high fields, EDMR
signals are dominated by the $\mathrm{P_b}$ transitions since
$\xi>1$ for both $\mathrm{^{31}P}-$\Pb and $\mathrm{P_b-P_b}$ pairs
while at the same time the $\mathrm{P_b-P_b}$ pair density is
significantly higher than the $\mathrm{^{31}P}-$\Pb pair density.


In conclusion, we have shown that EDMR on Si:P at the highest
magnetic fields reported to date allows us to observe the influence
of \Pb centers and $^{31}$P donor atoms on spin--dependent
photocurrents. In contrast to low magnetic field EDMR, there is no
intensity or transient correlation between these signals and, in
contrast to the \Pb signal, we find no evidence that the $^{31}$P
signals are due to interface processes. The intensity of the \Pb
signal increases linearly with temperature, vanishing as
$T\rightarrow 0$, which is shown to match the properties of charge
carrier tunneling between adjacent \Pb states. This effect is
expected only at high magnetic fields and does not contradict the
dominance of the well investigated $\mathrm{^{31}P}-$\Pb interface
recombination process at low magnetic field. Finally, we point out
that the spin-dependent tunneling demonstrated in this paper is
analogous to the mechanism proposed by Kane for readout of solid
state donor qubits \cite{kane}. Whilst previous attempts to
investigate this mechanism have relied on remote charge detection of
the transfer of an electron between clusters of donors
\cite{Buehler2006} and even two single donors\cite{Andresen2007},
this work demonstrates a spin dependent electronic tunneling
transitions between localized defect sites. Note that the
$\mathrm{^{31}P}-$\Pb mechanism that dominates at low magnetic
fields does not demonstrate this effect, as energy in not conserved.

This work was supported by a Visiting Scientist Program Grant
7300-100 from the National High Magnetic Field Laboratory. GWM was
supported by the EPSRC through grants GR/S23506 and EP/D049717/1.

\newpage
\newpage

\newpage

\end{document}